\documentclass[%
 reprint,superscriptaddress,
 amsmath,amssymb,
 aps,
 prmaterials
]{revtex4-2}

\usepackage{orcidlink}
\usepackage{textcomp}
\def\qe{\textsc{Quantum ESPRESSO}\texttrademark}
\usepackage{dcolumn}
\newcolumntype{d}[1]{D{.}{.}{#1}}
\usepackage{graphicx}
\usepackage{dcolumn}
\usepackage{bm}
\usepackage{xfrac}
\usepackage{nicefrac}
\usepackage{float}
\PassOptionsToPackage{unicode}{hyperref}
\usepackage{multirow, tabularray}
\usepackage{lineno}

\usepackage{xcolor}
\usepackage[normalem]{ulem}
\usepackage{lineno}

\definecolor{tangerine}{rgb}{0.944,0.522,0}
\definecolor{verde}{rgb}{0.,0.6,0}
\definecolor{rosso}{rgb}{0.9,0.0,0.2}
\definecolor{arancio}{rgb}{0.9,0.6,0.4}
\definecolor{viola}{rgb}{0.9,0.3,0.9}
\definecolor{turchese}{rgb}{0,0.5,0.5}

\newcommand{\editor}[2]{%
  \expandafter\newcommand\csname #1note\endcsname[1]{%
    \textcolor{#2}{(\textbf{#1:} ##1)}}%
  \expandafter\newcommand\csname #1\endcsname[1]{%
    \textcolor{#2}{##1}}%
  \expandafter\newcommand\csname #1cancel\endcsname[1]{%
    \textcolor{#2}{\sout{##1}}}%
  \expandafter\newcommand\csname #1change\endcsname[2]{%
    \textcolor{#2}{\sout{##1} ##2}}%
  \newenvironment{#1text}{\color{#2}}{\color{black}}
}

\editor{resub}{red}

\begin{document}

\preprint{APS/123-QED}

\title{
    Growing borophene on metal substrates: \\ a theoretical study of the role of oxygen on Al(111)
}

\author{Mandana Safari\,\orcidlink{0000-0003-2546-1594}}
\affiliation{SISSA -- Scuola Internazionale Superiore di Studi Avanzati, Trieste, Italy}
\author{Erik Vesselli}
\affiliation{Department of Physics, University of Trieste, Trieste, Italy}
\affiliation{CNR -- Istituto dell'Officina dei Materiali, TASC Laboratory, Trieste, Italy}
\author{Stefano de Gironcoli}
\affiliation{SISSA -- Scuola Internazionale Superiore di Studi Avanzati, Trieste, Italy}
\affiliation{CNR -- Istituto dell'Officina dei Materiali, SISSA unit, Trieste, Italy}
\author{Stefano Baroni\,\orcidlink{0000-0002-3508-6663}}%
\affiliation{SISSA -- Scuola Internazionale Superiore di Studi Avanzati, Trieste, Italy}
\affiliation{CNR -- Istituto dell'Officina dei Materiali, SISSA unit, Trieste, Italy}
\email{baroni@sissa.it}

\date{\today}

\begin{abstract}
    Charge transfer from a metal substrate stabilizes honeycomb borophene, whose electron deficit would otherwise spoil the hexagonal order of a $\pi$-bonded 2D atomic network. However, the coupling between the substrate and the boron overlayer may result in the formation of strong chemical bonds that would compromise the electronic properties of the overlayer. In this paper we present a theoretical study, based on state-of-the-art density-functional and genetic-optimization techniques, of the electronic and structural properties of borophene grown on Al(111), with emphasis on the impact of oxygen on the strength of the coupling between substrate and overlayer. While our results confirm the formation of Al-B bonds, they also predict that oxygen doping reduces charge transfer between aluminum and borophene, thus allowing modulation of their strength and paving the way to engineering the electronic properties of 2D-supported borophene sheets for industrial applications. Our study is completed by a thorough study of the thermodynamic stability of the oxygenated borophene-Al(111) interface.
\end{abstract}
\maketitle
\section{Introduction}\label{sec:intro}
2D boron allotropes are attracting increasing attention due to their promising electronic and chemical properties \cite{BBook,Erik2}. 2D metallicity, massless Dirac fermions \cite{rev2}, charge-density waves, superconductivity, and zero-tunneling barriers are some of the properties that make these materials extremely attractive for applications in the fields of quantum electronics, energy storage, catalysis, and sensoring \cite{BBook,Erik3}. Since the specific electron-phonon coupling characteristic of the (honeycomb) 2D B layers in borides accounts for the observed superconductive properties \cite{Erik4, SupconErik}, pure B materials would potentially pave the way to the design of boron-based 2D superconducting structures. In addition, interesting (photo)-catalytic properties of B-based monolayers have been reported for the hydrogen evolution reaction, a strategic reaction in the field of green hydrogen energy \cite{b2o}, and in the case of carbon dioxide reduction for the synthesis of energy vectors \cite{Erik6}. 

\begin{figure*}
     \centering
      \includegraphics[width=0.65\textwidth]{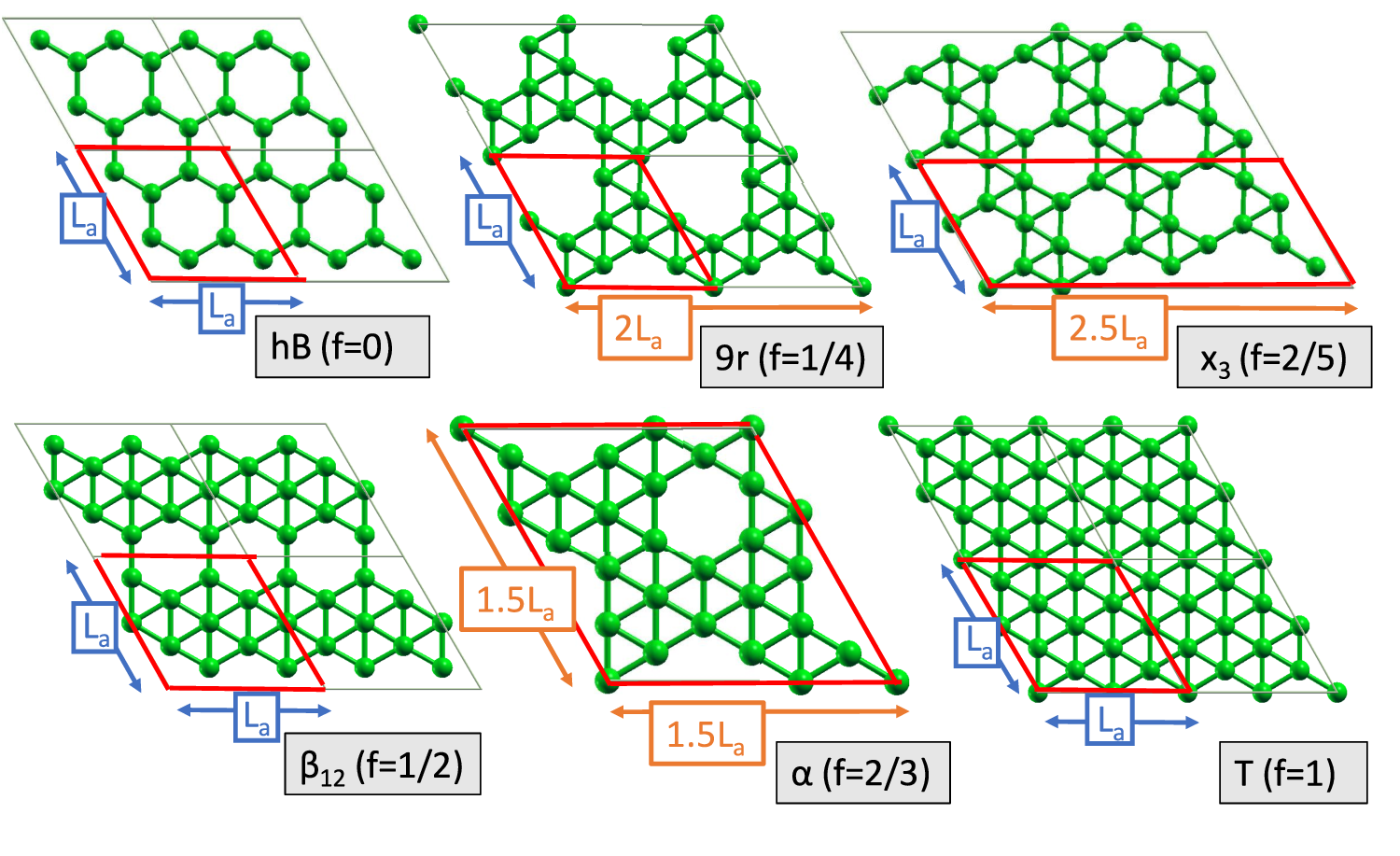} \caption{ \label{fig:fillfac} Representation of the honeycomb structure (hB) with zero filling factor (f $= 0$), $\rm {9r:f=\sfrac{1}{4}}$, $\rm {\chi_{3}:f=\sfrac{2}{5}}$, $\rm {\beta_{12}:f=\sfrac{1}{2}}$, $\rm {\alpha:f=\sfrac{2}{3}}$, and the triangular layer (T) with f $= 1$ (Scales are in units of L$_a$ = 5.71 \AA). The conventional cell of each polymorph is delimited with red parallelogram.}
\end{figure*}

Many of these premises are closely related to the electronic properties of a honeycomb 2D structure, whose stability in the case of borophene is however compromised by the electron deficiency of boron with respect to carbon. Ever since the first theoretical studies on borophene have appeared  \cite{rev1,rev3,BBook,3B9n}, it became apparent that this instability could be leveraged to tune its geometric, electronic, and chemical properties through self-doping, whereby the addition of an extra B atom to a honeycomb lattice releases three electrons to the planar skeleton \cite{Erik3}, thus helping stabilize it. The variety of structures that can be thus obtained is best described by introducing a \emph{filling parameter} (f), representing the fraction of hexagons hosting an extra boron atom \cite{B9n}. Correspondingly, the honeycomb structure (hB) is characterized by $\rm {f = 0}$, while the triangular layer (T) has $\rm {f = 1}$, representing the two self-doping extremes among a large family of structures ($\rm {9r:f=\sfrac{1}{4}}$, $\rm {\chi_{3}:f=\sfrac{2}{5}}$, $\rm {\beta_{12}:f=\sfrac{1}{2}}$, $\rm {\alpha:f=\sfrac{2}{3}}$) (see Fig. \ref{fig:fillfac}). Most of these polymorphs are metallic and have hexagonal or triangular structures (with the exception of the 9r structure, which displays nonagonal rings), while the bulk boron allotropes are generally semiconducting or insulating \cite{3BBook, 4BBook}. Amongst these structures, free-standing $\chi_{3}$ and $\beta_{12}$, were successfully synthesized \cite{freestanding}, while 9r and $\alpha$ are also considered as potentially stable without a support \cite{B9n}.

As mentioned above, the honeycomb structure, hB, attracts the most attention, because of its special features. A different, less disruptive, way of coping with electron deficiency is by charge transfer from a supporting metal substrate. Indeed, at variance with what is observed for Ag(111) \cite{ExpAg, ExpAg2}, honeycomb borophene (hB) has been successfully grown on an Al(111) termination, providing sufficient charge transfer \cite{ExpAl} to stabilize a flat, planar structure. The overlayer-substrate interaction resulting from this charge transfer, however, is so strong that an AlB$_2$ monolayer is unintentionally formed at the interface \cite{Erik10}. The question then naturally arises as to how the strength of the overlayer-substrate interaction can be tuned. Chemical doping of the interface can provide such a means, namely through oxidation, towards borophene oxide \cite{Erik4,boxide,Erik12}, or reduction, towards chemically stable borophane phases \cite{Erik13, Erik14, Erik15, Erik16}. In the former case, incorporation of atomic oxygen in the 2D crystal is mostly desirable with respect to oxygen bonding as a functional group, and this is largely expected in the case of borophene at variance with graphene \cite{boxide}. Various structures have been predicted in theory for both borophene and borophene oxides, although the experimental realization of supported borophene and its oxides appears quite challenging \cite{boxide}. As a step forward with respect to conventional atomic-layer deposition approaches for B, a possible novel route to the chemical synthesis of 2D B layers has been reported only very recently \cite{Erik17}.

In this work we present a thorough theoretical investigation that combines state-of-the-art electronic-structure methods based on density-functional theory (DFT) and novel genetic prediction approaches \cite{uspex,oganovB} to predict the impact of interface oxidation on the electronic properties and structural stability of hB grown on Al(111). Our extensive search for the stable structures of B$_{x}$O$_{y}$/Al(111) allows us to draw some conclusions on the phase diagram of this system. Depending on the oxygen chemical potential, we find that Al atoms are extracted from the surface and incorporated into the hB phase that becomes progressively rougher as a function of the oxygen concentration. A larger charge transfer from the substrate to the B layer across the interface does not necessarily correspond to an increased stability and decoupling from the substrate, while we observe that a sizeable intralayer charge redistribution also plays a role, in association with the buckling induced by the distortion of bonding orbitals.

The rest of our paper is organized as follows: in Sec. \ref{sec:AlB} we investigate boron polymorphs syntesized on Al(111): structure prediction of pristine borophene is presented in Sec. \ref{sec:structAlB}, followed by a discussion of charge-transfer mechanisms in Sec. \ref{sec:chAlB}. The effects of oxygen doping are presented in Sec. \ref{sec:AlBoxygen}: we first introduce the predicted structures in Sec. \ref{sec:structO}, and then describe charge-transfer mechanisms in Sec. \ref{sec:chO}. In Sec. \ref{sec:thermo} we present the phase diagram of the oxidized B{@}Al(111) interface. Finally, our results are summarized in Sec. \ref{sec:conc}. 

\section{Borophene on A\lowercase{l} (111)}\label{sec:AlB}

\begin{figure}
\includegraphics[width=0.45\textwidth]{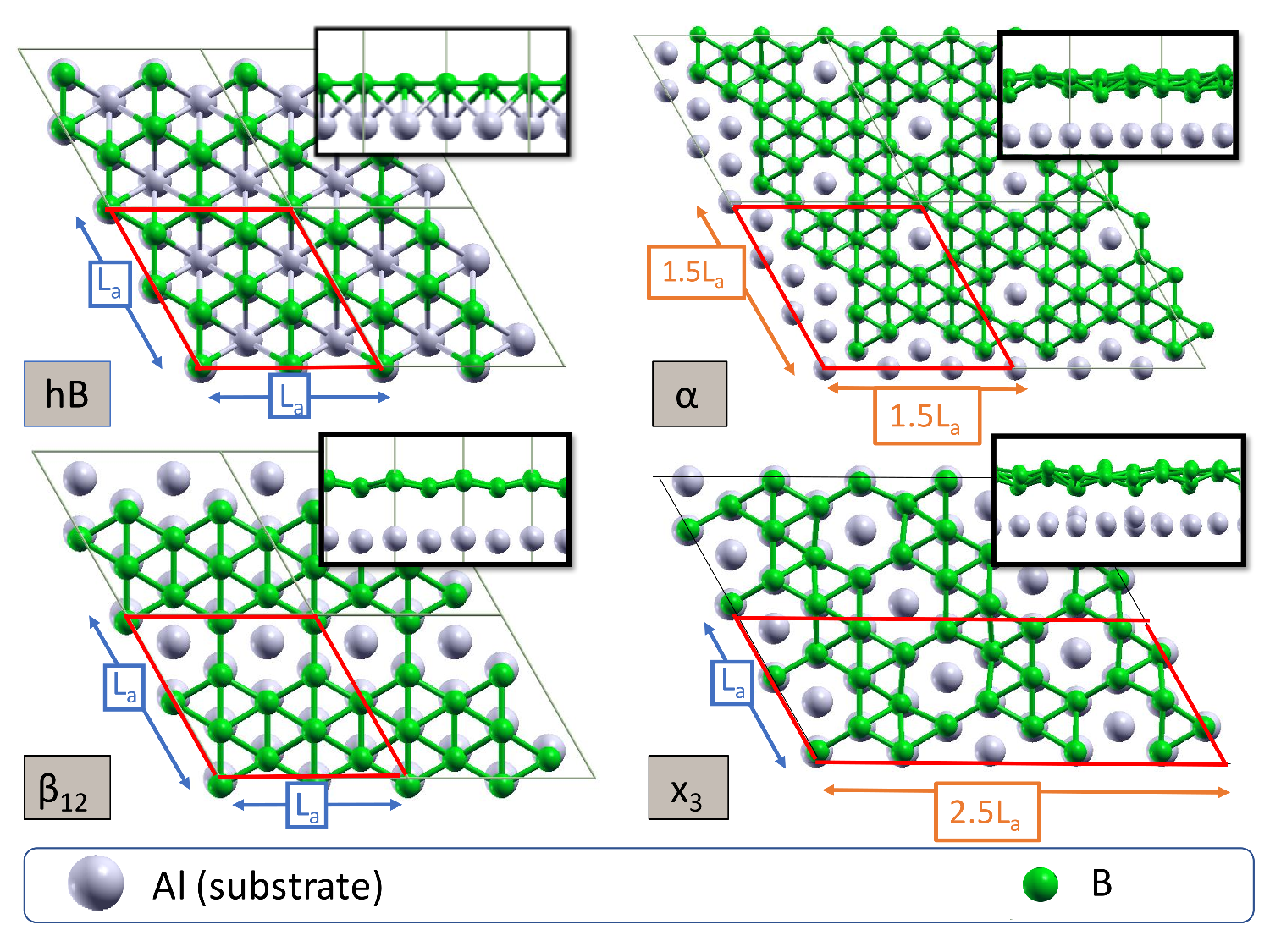}
\caption{\label{fig:AlB-p}
Geometrical structure of different borophene polymorphs on Al(111). The double arrows indicate the size of the unit cells, in units of $\mathrm{L_a} = 5.71$,$\mathrm{\text{\AA}}$. The conventional cell of each polymorph is delimited with red parallelogram.}
\end{figure}

Al(111) provides a lattice-matched substrate, as well as an electron-charge reservoir, for the growth of the borophene polymorphs examined in the introduction, with the potential of stabilizing the sought-after honeycomb one, H (see Fig. \ref{fig:AlB-p} (a)) \cite{ExpAl}.

\begin{table*}
      \caption{Filling factors$^a$ and surface energies, $e_a$, of various borophene polymorphs grown on Al(111), see text. }
    \label{tab:1} 
    \begin{ruledtabular}
    \begin{tabular}{ l | d{2.4} d{2.4} d{2.4} d{2.4} d{2.4} d{2.4} d{2.4} d{2.4}}
       & \multicolumn{1}{c}{hB} &  \multicolumn{1}{c}{$\alpha$} & \multicolumn{1}{c}{$\beta_{12}$} & \multicolumn{1}{c}{$\chi_3$} & \multicolumn{1}{c}{U$_1$} & \multicolumn{1}{c}{U$_2$} &   \multicolumn{1}{c}{U$_3$} & \multicolumn{1}{c}{U$_4$}  \\
        \hline
        Filling factor\footnote{Number of filled rings of of borophene (visualized in fig. \ref{fig:fillfac})} &  0 & 0.666 & 0.5 & 0.4 & - & - & -  & - \\ 
        Boron coverage\footnote{Ratio of the number of atoms in the B overlayer, to the number of Al atoms in the first substrate layer.
        } &  2 & 2.666 & 2.5 & 2.4 & 2 & 2 & 2  & 2 \\ 
         e$_{a}$ \big[ $\mathrm{eV}$/$\mathrm{\text{\AA}}^2$ \big] &  0.086 & 0.114 & 0.096 & 0.093  & 0.104 & 0.113  & 0.112 & 0.097 \\
    \end{tabular}
    \end{ruledtabular}
\end{table*}

\subsection{Structure Prediction of Borophene on Al(111) \label{sec:structAlB}}
\begin{figure}
    \centering
    \includegraphics[width=0.45\textwidth]{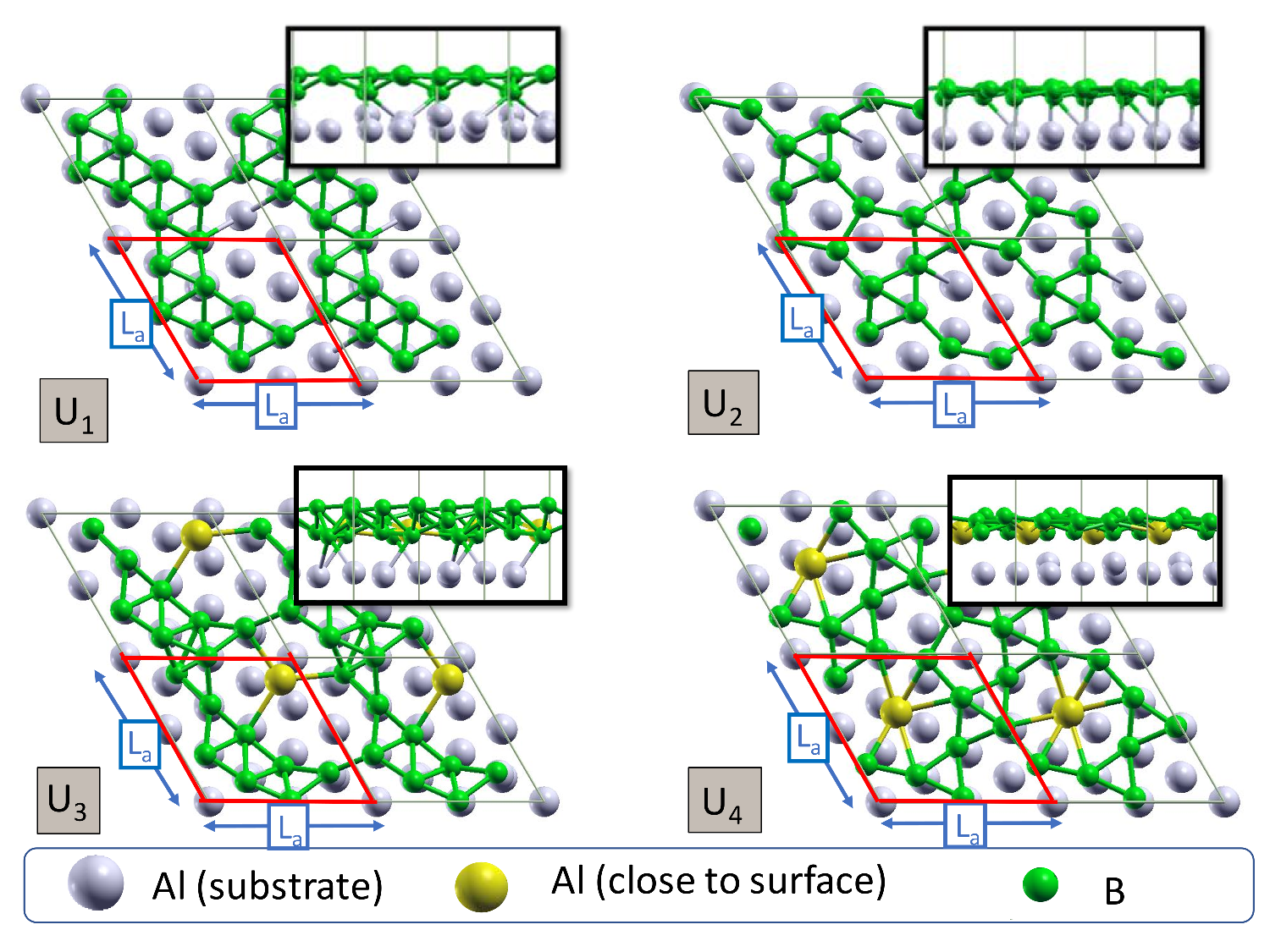}
     \caption{Top and side view of U$_1$, U$_2$, U$_3$, and U$_4$ polymorphs of borophene. The gold color of aluminum is considered to show the aluminum atoms trapped in surface rings, or separate from substrate and tend to adsorb by surface because of higher electronegativity of boron (the length scale is $\mathrm{ L_a = 5.71~\text{\AA}}$). Conventional cell of each polymorph is determined with red hexagon.}\label{fig:AlB-uu}
\end{figure}

In order to find the most stable structures of borophene on Al(111), we have computed the surface energies, $E_s$, of the polymorphs shown in Fig. \ref{fig:AlB-p}, defined as:
\begin{multline} \label{eq:e_a}
     \quad e_{a}=\Bigl ( E_a\bigl [\mathrm{B@Al}(111)\bigr ]
     \\ - E\bigl [ \mathrm{Al}(111)\bigr ]-N_\mathrm{B}\times E_b\bigl [\mathrm B\bigr ]
    \Bigr ) \Big / S, \quad
\end{multline}
where the label `$a$' indicates the polymorph, $E_b[\mathrm B]$ is the energy per atom of a B bulk crystal, and $N$ the number of B atoms contained in the area $S$ of the surface. We then focused on the stability of the honeycomb structure with respect to other allotropes with the same stoichiometry, and ran a thorough structural search using a genetic algorithm (an overview of the computational details is presented in Appendix \ref{app:comp}). A summary of our results is presented in Table \ref{tab:1}, including four newly discovered polymorphs (dubbed U$_n$, with $n=1,\cdots 4$, see Fig. \ref{fig:AlB-uu}). One of the defining features of these new polymorphs is the presence of a variety of polygonal rings (pentagons, octagons, and decagons), besides the triangular, hexagonal and nonagonal stable configurations. This fact proves that boron is flexible enough to accept a wide variety of combinations of bonding electrons as well. The surface energies defined in Table \ref{tab:1} equal the static contributions to the surface free energies \cite{scheffler1st, initTR}, when the B chemical potential equals the value in the bulk, \emph{i.e.} at growth $pT$ conditions where bulk B would be in equilibrium with its own atomic vapor. The positive values of these energies show that the formation of borophene on Al(111) is endothermic; a comparison of the various surface energies indicates that the hB phase would be favored at these conditions. More on the stability of various borophene phases on Al(111) at varying growth conditions in Sec. \ref{sec:thermo}.

\subsection{Charge-transfer mechanisms}\label{sec:chAlB}

\begin{figure}[h]
   \includegraphics[width=0.45\textwidth]{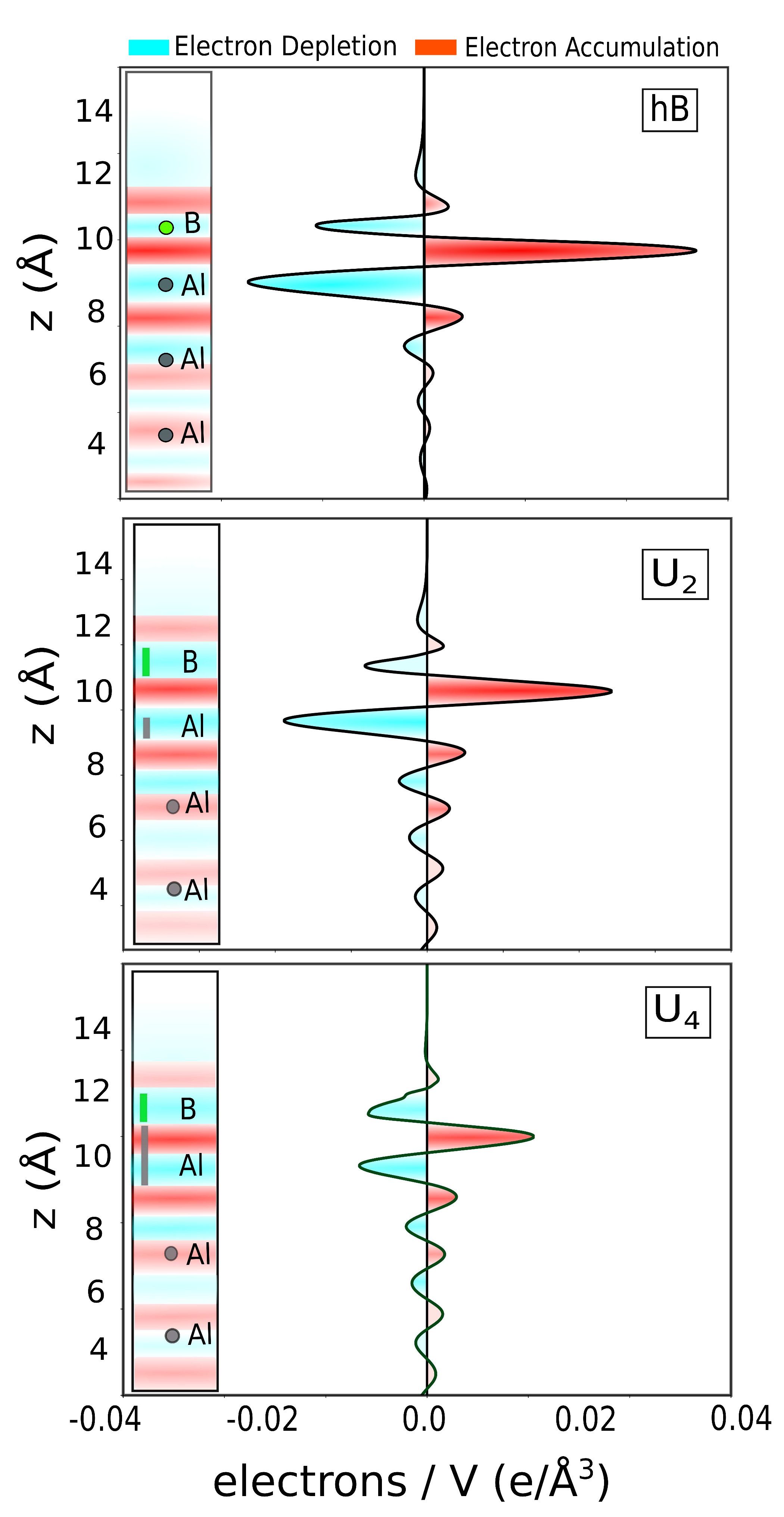}
   \caption{\label{fig:ch-compare} Charge density differences between aluminum and borophene in honeycomb phase, U$_2$, and U$_4$, the intensity of color associated with higher values, red and cyan denote electron accumulation depletion, respectively. The vertical bars next to the chemical symbols of the various elements indicate the spread in the vertical positions of the corresponding atoms. }
\end{figure}

In order to figure out the charge transfer processes occurring at the B/Al interface and their effects on structural stability and experimental findings, we consider the planar average of the displaced charge distribution of the system, defined as:
\begin{equation} \label{eqn:drho}
   \Delta \rho(z) = \frac{1}{S} \int_S \bigl (\rho_{AlB}(\bm r) - \rho_{Al}(\bm r) - \rho_{B}(\bm r) \bigr )dxdy ,
\end{equation}
where $\rho_{AlB}$ denotes the electron charge-density of the B{@}Al(111) system, $\rho_{Al}$ and $\rho_{B}$ those of the two constituents with their atoms clamped at the positions they would have at the interface, and the integral is performed over planes perpendicular to the growth direction ($z$). In Fig. \ref{fig:ch-compare} we display the planar average of the displaced charge-density distribution, Eq. \ref{eqn:drho}, for the honeycomb, U$_2$, and U$_4$ structures. This charge-displacement pattern can be qualititavely described as resulting from the balance of two effects. The first is the charge transfer between the aluminum substrate and the more electronegative boron overlayer. The second effect is the formation of chemical bonds between the substrate and the overlayer, which determine a charge accumulation in-between at the expense of charge depletion in the atomic layers. The second effect is larger than the first, always resulting in a strong charge-accumulation peak in the bonding region. The first effect determines an asymmetry in the charge depletion in the two atomic layers, which is stronger in aluminum than in boron. This asymmetry is greatest in the hB structure and almost vanishing in $\mathrm U_4$---as apparent in the figure and from the integrated  charges reported in Table \ref{tab:0}---thus confirming the stabilizing effect of charge transfer on the hB structure.\\
\begin{table}[h]
\caption{\label{tab:0} Charge transfer between the Boron and outer Aluminun layers and the bond region in-between for different B$@$Al(111) polymorphs, as obtained by integrating the planar averages of the charge-density differences displayed in Fig. \ref{fig:ch-compare} over the corresponding peaks. Units are $\text{milli-electrons}/\mathrm{\text{\AA}}^2$.}
\begin{ruledtabular}
\begin{tabular}{l| d{2.4} | d{2.4}  | d{2.4} }
                            & hB & U_2 & U_4 \\
        \hline
    Bond        & +20.8 & +14.5 & +8.2 \\
    Boron       &  -6.1 &  -3.3 & -4.9 \\     
    Aluminum    &  -16.9 & -12.0 & -5.3
\end{tabular}
\end{ruledtabular}
\end{table}
\\
\section{Oxidation of the B/A\lowercase{l} (111) interface \label{sec:AlBoxygen}}
Oxygen is a common, highly reactive, gas and its impact on surface morphology and interface formation can be significant upon dissociative chemisorption. Oxygen can affect the structure of the substrate and of the overlayer, as in ZnO \cite{CdXZnX} and B$_2$O \cite{b2o}, or may have disruptive effects, as in the corrosion process in iron. The question then naturally arises as to which effects the presence of oxygen may have on borophene formation, and on their dependence on the details of oxygen surface incorporation.

\subsection{Structural prediction in the presence of oxygen\label{sec:structO}}

\begin{figure*}
       \includegraphics[width=0.8\textwidth]{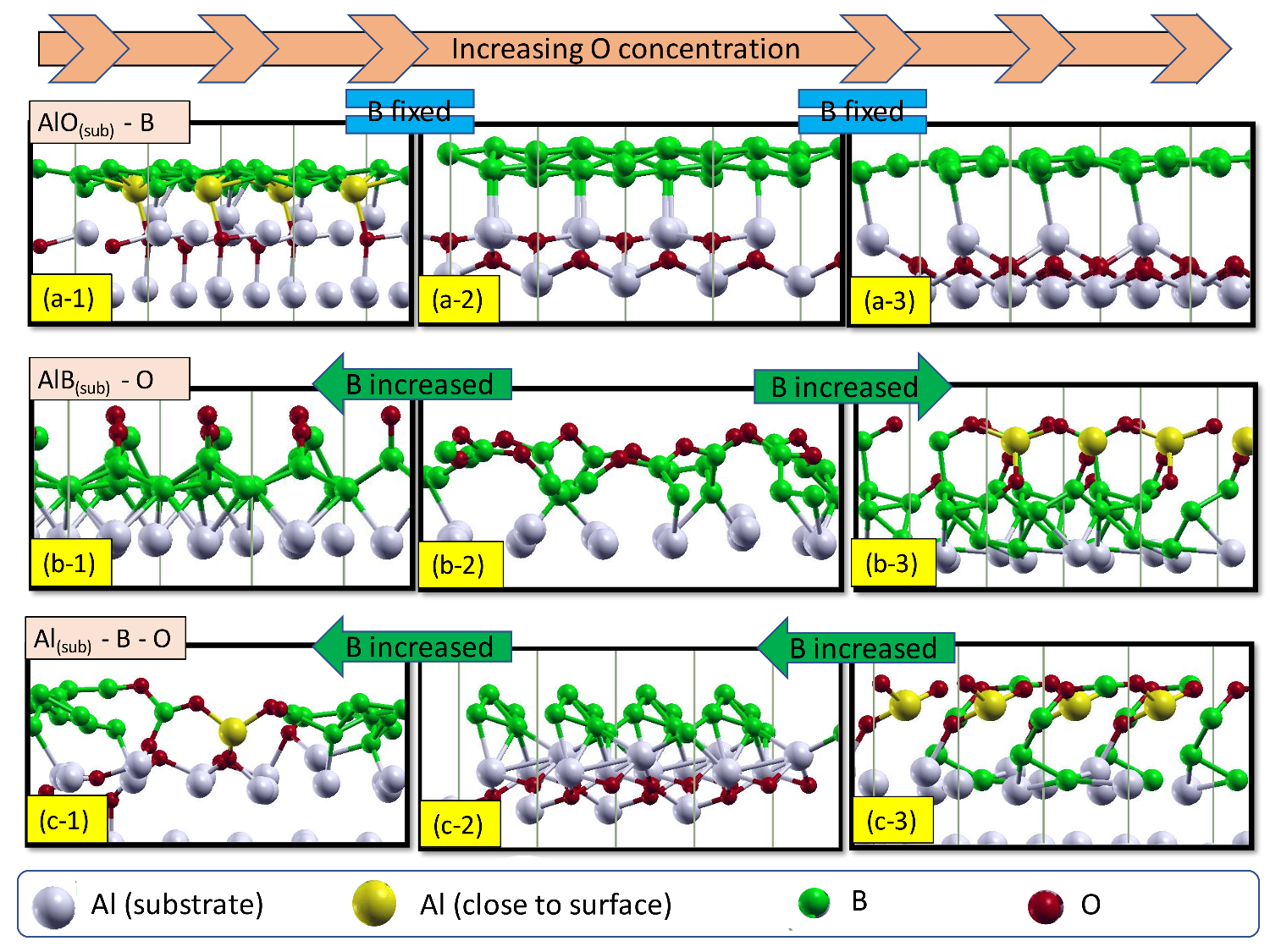}
       \caption{\label{fig:p-all} Three different setups of structure prediction. (a) Structures with AlO$_x$ substrate (fixed B coverage 2 ML), with oxygen coverages of \sfrac{1}{2} ML (a-1), \sfrac{3}{4} ML  (a-2), and ML (a-3). (b) Structures with AlB$_x$ substrate, with oxygen coverage of: \sfrac{1}{2} ML (b-1), \sfrac{2}{3} ML (b-2), 1 ML (b-3) (boron coverags: 2, \sfrac{4}{3}, and 2 ML respectively). (c) Structures with Al substrate, with oxygen coverage of: \sfrac{2}{3} ML (c-1), \sfrac{3}{4} ML (c-2), 1 ML (c-3) (boron coverages: \sfrac{4}{3}, \sfrac{5}{4}, and 1 ML respectively). Aluminum atoms completely extracted from the substrate and incorporated in the overlayer are shown in yellow.}
\end{figure*}

In order to study the tendency of oxygen and borophene to combine at an aluminum surface, we perform several genetic structural optimizations starting from different initial configurations mimicking different experimental conditions. In the following, we will sometimes refer to each one of these initial configurations as to a \emph{setup}. In all cases we seek the lowest-energy configuration accessible from a given setup for a given stoichiometry, without, at this stage, accounting for any thermal or kinetic effects, such as temperature or the different partial pressures of the elements being adsorbed from the gas phase. Three qualitatively different setups are considered in the structural exploration, leading to the structures illustrated in Fig. \ref{fig:p-all}. The surface coverage, expressed in monolayer (ML), is defined here as the ratio between the number of adsorbed atoms (boron, oxygen) with respect to the number of Al atoms of the terminal substrate layer (in initial configuration) per unit cell.

First, we suppose the aluminum surface to be passivated by pre-exposure to oxygen, corresponding to different O:Al  re-coverages (a-1: \sfrac{1}{2} ML; a-2: \sfrac{3}{4} ML, and a-3: 1 ML). Boron atoms are then allowed to form the most stable configuration on top of these oxidized substrates (we refer the reader to Appendix \ref{app:comp} for details on the calculations). This setup mimicks the effects of oxygen passivation of the aluminum substrate on borophene formation, and also sheds light onto the possible formation of honeycomb structure on aluminum oxide. The trend with respect to increasing oxygen concentration is clear: with higher oxygen coverage the topmost aluminum atoms tend to bind preferably to oxygen and boron atoms are increasingly decoupled from the oxidized substrate. However they tend to form triangular bonds, forming $\alpha$, $\chi_3$, and $\beta_{12}$ structures rather than the honeycomb polymorph. As mentioned earlier, although these structures can be considered as energetically stable, the polymorph actually observed in an experiment will depend on the specific pressure/temperature growth conditions, as discussed in Sec. \ref{sec:thermo}.

In a second setup, we investigate the stability of a pre-formed borophene overlayer to oxygen exposure. In this case, (second row of Fig.~\ref{fig:p-all}), boron overlayers on Al(111) with two surface ratios, B:Al = 2 for b-1 and b-3, and B:Al = \sfrac{4}{3} for b-2, are exposed to an increasing amount of atomic oxygen (with O:Al ratios ranging from \sfrac{1}{2} in b-1, to \sfrac{2}{3} in b-2, to 1 in b-3). Upon oxygen adsorption, a significant disruption of the borophene overlayer occurs as the highly electronegative oxygen tries to catch electrons as much as possible from its neighbors. The borophene layer reorganizes to the point that, at high oxygen concentration, channels are created through which aluminum atoms close to the borophene layer, that are already observed in some low energy borophene-on-Al structures even in absence of oxygen adsorption (see section \ref{sec:structAlB} and Fig.~\ref{fig:AlB-uu}), migrate to the surface and directly bind to oxygen.

Finally, simultaneous adsorption of both oxygen and boron is considered, with different relative concentrations for a fixed combined (B+O) coverage of 2 ML (O:Al = \sfrac{2}{3} and B:Al = \sfrac{4}{3} in c-1, O:Al = \sfrac{3}{4} and B:Al = \sfrac{5}{4} in c-2, and O:Al = 1  and B:Al = 1 in c-3). In this setup the two adsorbed species compete for bonding with the substrate atoms. This case is also designed to detect the possible  creation of a boron oxide overlayer on the aluminum substrate \cite{boxide}. Our results dismiss this possibility: when in excess, boron tends to cluster while oxygen binds preferably to aluminum forming an aluminum oxide layer. Occasionally, some boron-oxygen bond is observed but clearly boron atoms prefer to attach to aluminum or cluster on their own. This trend makes us conclude that borophene resists against oxidation, at least in the presence of aluminum.

Overall, our simulations indicate that oxygen preferentially binds to aluminum rather than to boron, except when access to the substrate atoms is hindered by the pre-deposition of a boron layer. Even in this case the interaction with the strongly electronegative oxygen atoms tends to disrupt the borophene overlayer that would need electron donation to be stabilized in a flat configuration. No tendency toward the formation of a boron oxide layer is observed. Boron atoms prefer instead to cluster on their own, mostly forming triangular bonds, or bonding to aluminum atoms.

\subsection{Charge transfer in the presence of oxygen \label{sec:chO}} 

\begin{figure}[h]
      \includegraphics[width=0.45\textwidth]{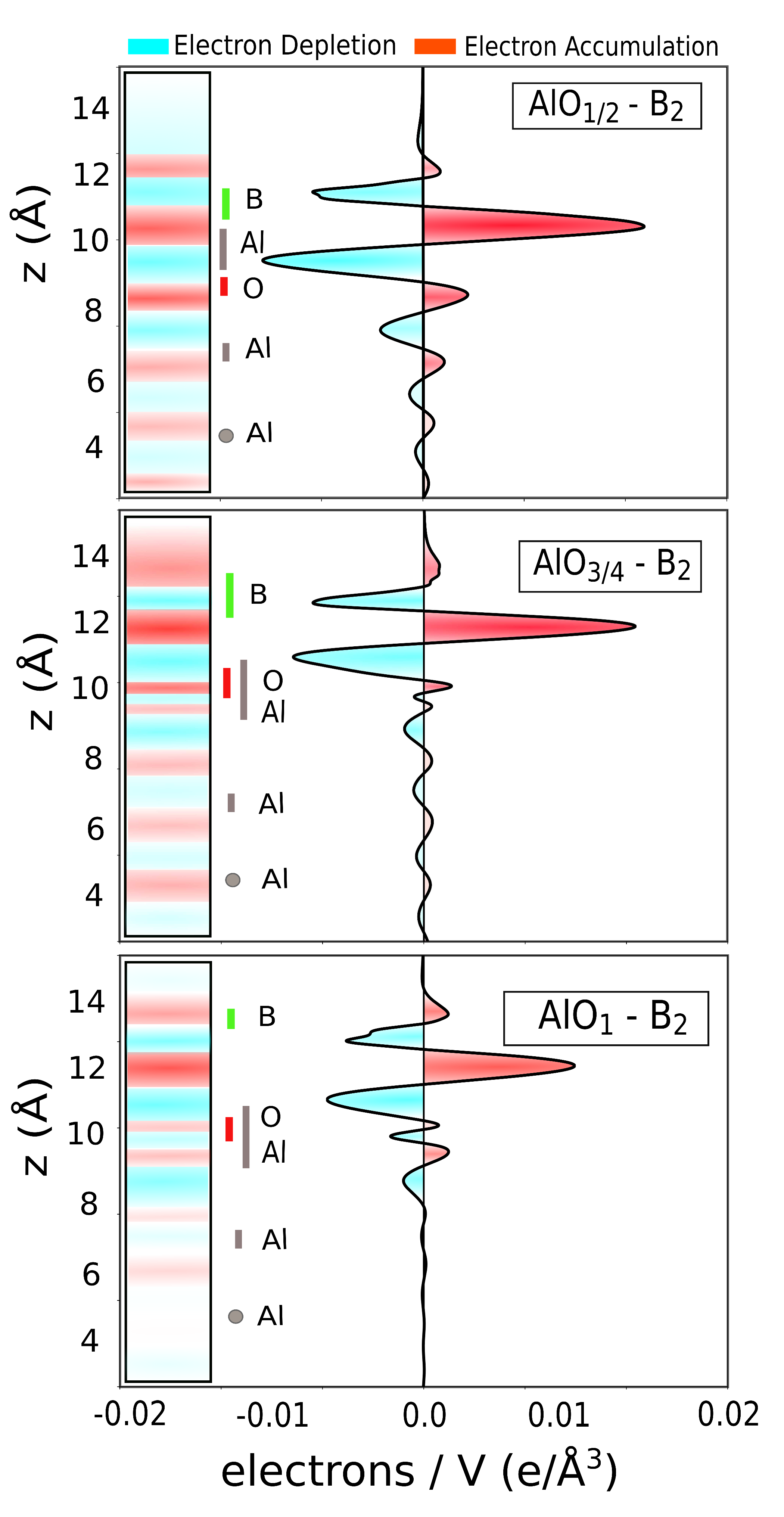}
      \caption{\label{fig:ch-alo} Charge density differences between aluminum and borophene in the presence of Oxygen in first setup (AlO$_x$ - B$_2$), and second setup (AlB$_x$ - O$_y$). Green span denotes the boron position, and red span shows the oxygen location of each structure. The vertical bars next to the chemical symbols of the various elements indicate the spread in the vertical positions of the corresponding atoms.}
\end{figure}

As seen in the previous section, a different sequence in the surface exposure to boron and oxygen leads to different final configurations. This can be further analyzed by examining the corresponding charge-transfer profiles. Charge depletion and accumulation for the three structures from the first row of Fig. \ref{fig:p-all} are displayed in Fig. \ref{fig:ch-alo}, which shows the planar averages of the charge density difference between the full systems and the superposition of the individual components of overlayer (B and O) and of the Al substrate (see equation \ref{eqn:drho}). The main feature resulting from this figure is the accumulation of bonding charge between the overlayer and the substrate, counterbalanced by a depletion on Al and B layers. The charge density depletion around Al atoms is enhanced with increasing oxygen concentration.

Total-energy comparison of similar structures, such as AlO$_{\sfrac{1}{2}}$--B$_2$ (Fig.~\ref{fig:p-all}, panel a-1) and AlB$_2$--O$_{\sfrac{1}{2}}$ (b-1) or AlB$_{\sfrac{4}{3}}$--O$_{\sfrac{2}{3}}$ (b-2) and Al--B$_{\sfrac{4}{3}}$--O$_{\sfrac{2}{3}}$ (c-1), suggests that oxygen tends to bind to Al atoms, rather than to boron, a fact that could in principle be exploited to reduce the strength of the B--Al interaction. This is confirmed by L\"owdin charge population analysis in Table \ref{tab:3}, where we report the deviation from the atomic nominal valence charge. The order of the structures listed in the table follows the one used in Fig. \ref{fig:p-all}. We can follow the trend of charge transfer across the table. For systems where oxygen is pre-adsorbed on the Al substrate (first row of figure \ref{fig:p-all}), a strong transfer from aluminum to oxygen is observed, consistently with the large difference in electronegativity between the two species. Some donation from Al to B is also observed for low oxygen concentration, which is blocked as the oxygen fraction increases, and the borophene layer becomes more detached.

\begin{table*}
\caption{\label{tab:3}The average values of accumulated and depleted electron per atom, calculated based on L\"owdin population analysis, in different structures. For some structures the average L\"owdin charge of Al atoms is further decomposed in the contribution from Al atoms strongly bound to atoms in the overlayer (B and/or O depending on the structure), marked as yellow in Fig.~\ref{fig:p-all}, and the rest of the Al atoms, light-gray.}
\begin{ruledtabular}
\begin{tabular}{c| d{2.4} d{2.4} d{2.4} | d{2.4} d{2.4} d{2.4} | d{2.4} d{2.4} d{2.4} }
                        & \multicolumn{1}{c}{(a-1)} & \multicolumn{1}{c}{(a-2)}  & \multicolumn{1}{c}{(a-3)} & \multicolumn{1}{c}{(b-1)} & \multicolumn{1}{c}{(b-2)} & \multicolumn{1}{c}{(b-3)} & \multicolumn{1}{c}{(c-1)} & \multicolumn{1}{c}{(c-2)} & \multicolumn{1}{c}{(c-3)}  \\
        \hline
      L\"owdin O (avg./atom) & 0.767 & 0.804 & 0.808 & 0.330 & 0.341  & 0.538 & 0.650 & 0.809 & 0.554   \\
        
      L\"owdin B (avg./atom) & 0.225 & 0.037  & 0.002 & 0.065 & -0.049 & 0.047 & -0.002 & 0.228 &  0.013   \\
    \hline
      L\"owdin Al (avg./atom) & -0.875 & -0.874 & -1.061 & -0.446 & -0.268 & -0.748 & -0.536  & -1.034 & -0.650 \\
     L\"owdin Al-yellow (avg./atom) & -1.231 & ... &  ... & ... & ... & -1.490 & -1.137  & .... & -1.313 \\
     L\"owdin Al-grey (avg./atom) & -0.757 & -0.874 & -1.061 & -0.446 & -0.268 & -0.501 & -0.462  & -1.034 & -0.430 \\
\end{tabular}
\end{ruledtabular}
\end{table*}

In the second setup (second row of Fig. \ref{fig:p-all}) oxygen is adsorbed on top of an Al--B system and a strong reduction in the charge transfer can be observed. Again, Al atoms lose electrons to oxygen while B atoms remain largely unaffected merely shuttling the charge from the metallic substrate to oxygen. As the oxygen content increases and oxygen atoms find ways to directly bind to Al (figure \ref{fig:p-all} panel b-3) the Al-O charge transfer becomes significant again.

Finally, if O and B atoms are adsorbed at the same time in different concentrations (third row of Fig. \ref{fig:p-all}, panels c-1 to c-3), B atoms prefer to form clusters with triangular bonds rather than competing with oxygen to bind to aluminum. Strong Al-O charge transfer results across the direct Al-O bonds while some Al-B charge transfer occurs only when B is directly in contact with aluminum. Aluminum is clearly the less electronegative of the three species, always loosing electrons to its neighbors, especially when forming close bonds with the atoms in the overlayer (yellow atoms in fig.~\ref{fig:p-all}). These findings confirm that---because of the reduced charge transfer from aluminum to borophene due to the competing adsorption of the more electronegative oxygen---charge-doping of borophene is less likely to occur. Stabilization of the borophene layer occurs therefore through self-doping and the hB borophene polymorph has less chances to form.

\section{Phase diagram from ab-initio thermodynamics \label{sec:thermo}}

All of the considerations made so far are based on purely energetic arguments and only apply to isolated systems, whose total energy and number of atoms of each chemical species are held fixed. The actual structures observed in an experiment result from a combination of kinetic and thermodynamic effects. While the former are very difficult to model, they are nevertheless driven by the tendency towards thermodynamic equilibrium between the sample and its environment (essentially, the vapour present in the growth chamber), with which it can exchange atoms and heat. In order to account for these effects, we make use of concepts from \emph{ab initio thermodynamics}, a methodology made popular by Scheffler and co-workers in the late nineties \cite{initTR,ab-init-2}.

In a nutshell, the thermodynamically stable structure is the one that---for given values of temperature, $T$, and chemical potential of each atomic species, $\mu_X$---minimizes the surface (grand) free energy per unit area. At equilibrium, the chemical potentials of the various elements at the surface should equal those in the vapour phase. Assuming an ideal-gas law, which always holds at low pressure, the chemical potential of the atomic species $X$ reads: $\mu_X(p,T) = \mu^\circ(T) +k_BT\log(p/p^\circ)$, where $p^\circ$ is the pressure at \emph{standard conditions} (\emph{e.g.} the pressure at which the vapour is in equilibrium with the bulk or some specific surface structure at a given temperature) and $\mu^\circ(T)=\mu(p^\circ,T)$ the corresponding chemical potential. A commonly adopted approximation is to neglect the vibrational contribution to the surface free energy, so that \emph{e.g.} the surface grand free energy of polymorph $a$ in the absence of oxygen reads:
\begin{equation} \label{eq:surf-grand}
    \varphi_a(\mu_{B}) = e_a -n_B\mu_{B},
\end{equation}
where $e_a$ is the polymorph's surface energy per unit area, Eq. \eqref{eq:e_a}, $n_B$ the number of B atoms per unit area, and the B chemical potential is referred to the bulk, \emph{i.e.} the B vapour in the growth chamber is supposed to be in equilibrium with its own bulk. The reference data for the B chemical potential entering Eq. \eqref{eq:surf-grand} are taken from Ref. \onlinecite{NIST}.

\begin{figure}[h]
       \includegraphics[width=0.48\textwidth]{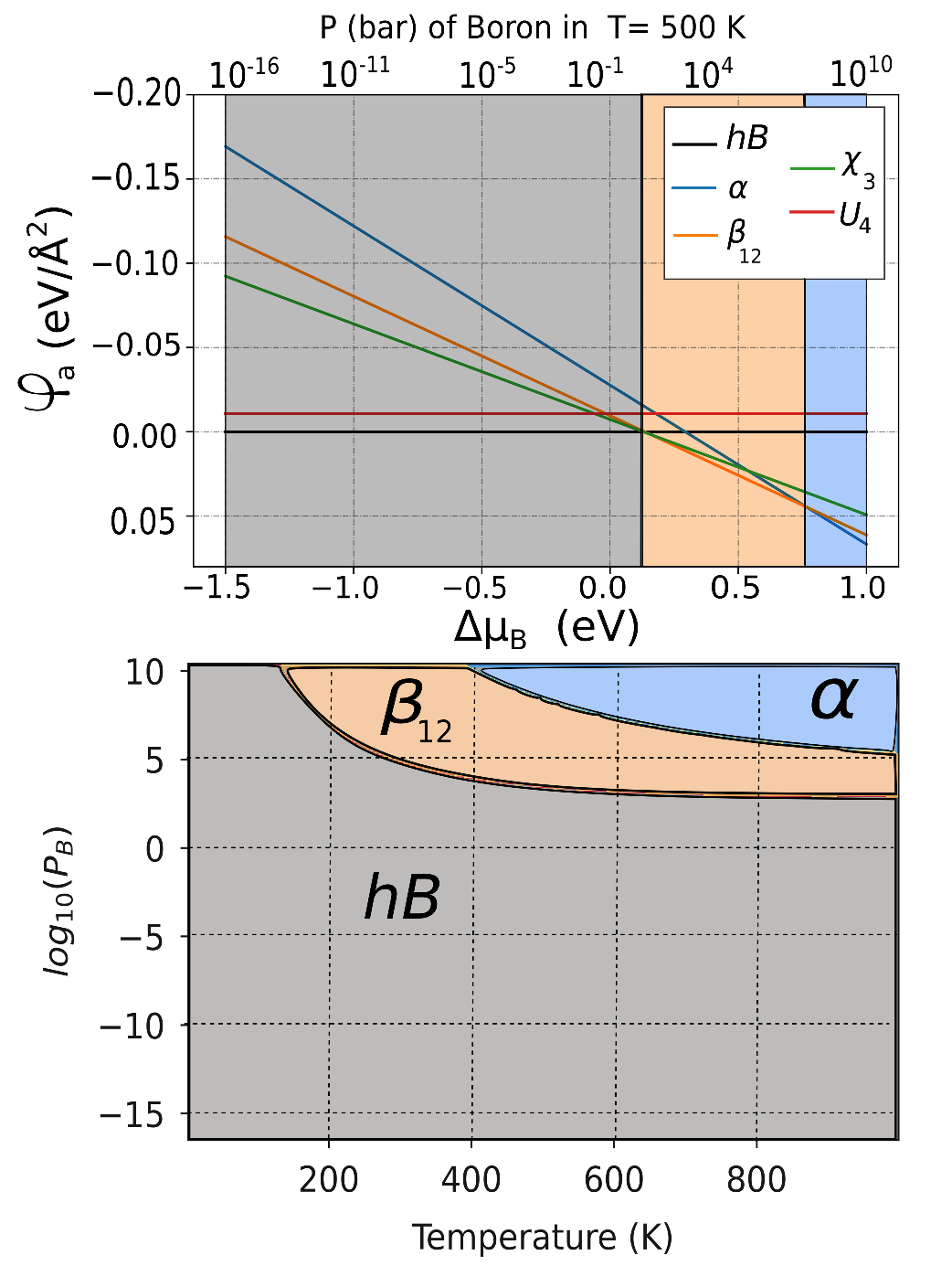}
       \caption{\label{fig:PhaseB} Upper panel: dependence of the surface grand free energy (Eq. \eqref{eq:surf-grand}) on the B chemical potential, for various borophene allotropes grown on Al(111). The hB surface energy and the B chemical potential of the bulk are taken as references. Lower panel: phase diagram of borophene on Al(111). Temperature is expressed in Kelvin and pressure in bar
       .}
\end{figure}

In the upper panel of Fig. \ref{fig:PhaseB} we display the adsorption free energy per unit area, Eq. \eqref{eq:surf-grand}, of various borophene polymorphs grown on Al(111), as a function of the B chemical potential. Energies are referred to that of the hB allotrope, whereas the B chemical potential is referred to its bulk value. According to these data, the hB polymorph is dominant at low chemical potential. Then, at intermediate chemical potential ($\mu_B \simeq 0.13$ to $0.77~\text{eV}$), the $\rm {\beta_{12}}$ phase becomes dominant, whereas the $\rm {\alpha}$ structure of B@Al(111) is formed when the B concentration further increases, \emph{e.g.} at even higher chemical potential. The lower panel of Fig. \ref{fig:PhaseB} reports the P$-$T phase diagram obtained with the method used by Molinari \emph{et al.} \cite{Molinari} and implemented in the \texttt{Surfinpy} package \cite{surfinpy}. Since the phase diagram only shows the most stable structures, other local minima in configuration space have no chance to show off, but their relative stability is relevant to understand the mechanisms determining the structures observed experimentally, and thus worth studying. The lower panel of Fig. \ref{fig:PhaseB} summarizes these findings in a phase diagram with respect to pressure and temperature. The lower panel reports the phase diagram of borophene on Al(111) deduced from these data. While the relevance of these data to the growth of B on Al(111) is limited by both the overall accuracy of the theoretical calculations and by the actual growth conditions \cite{growth}---which are certainly far from thermodynamical equilibrium---our results indicate that at low temperature and pressure there exists a window of thermodynamical stability of the hB phase.

\begin{figure}
        \includegraphics[width=0.5\textwidth]{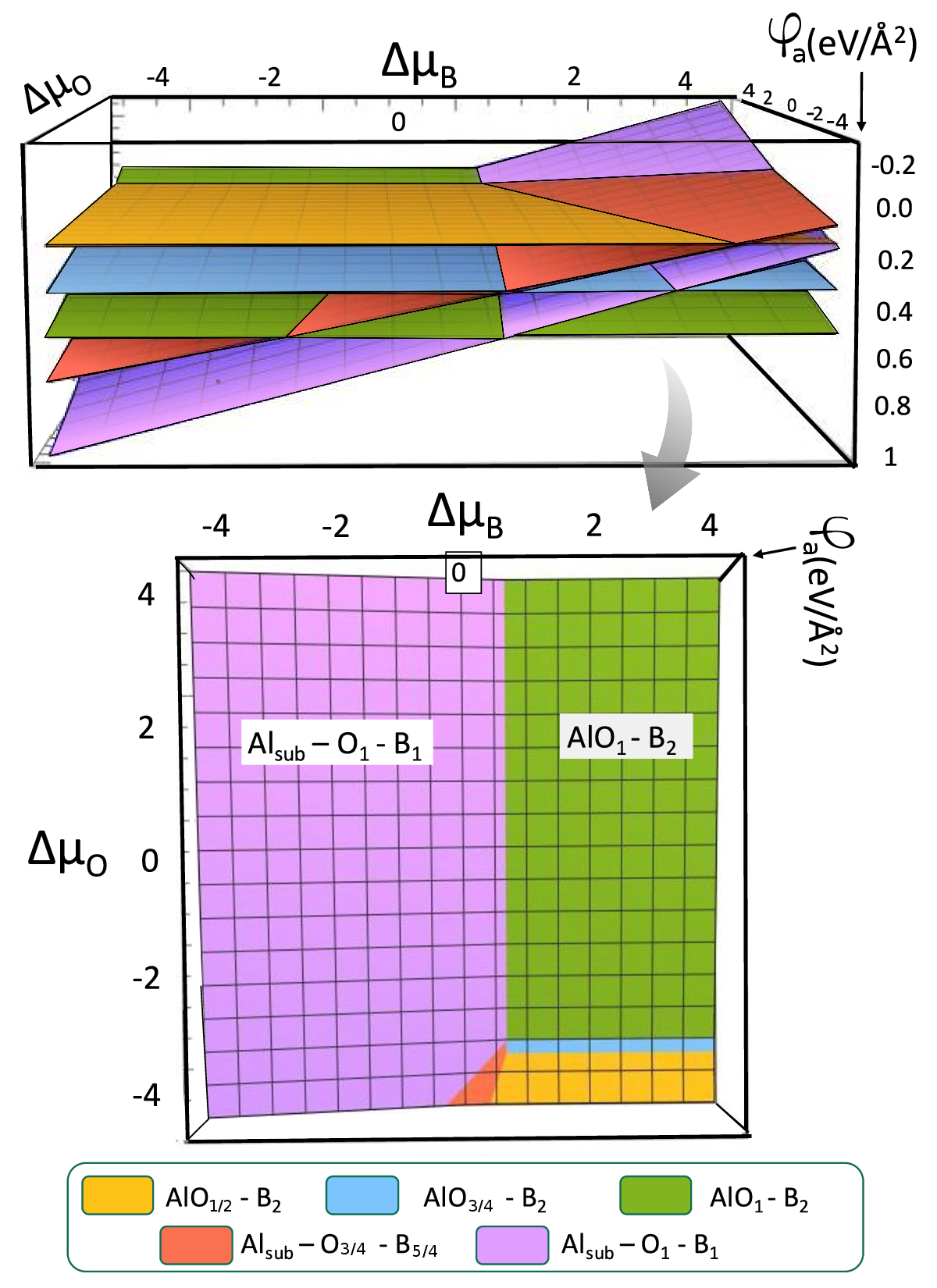}
        \caption{\label{fig:3Dphase} 3D phase diagram, showing the adsorption Gibbs free energy of the system as a function of B and O chemical potentials.}
\end{figure}

These considerations can be easily extended to an analysis of the stability of B@Al[111] in the presence of oxygen. Our results are reported in Fig.  \ref{fig:3Dphase}, which displays the adsorption free energies of various structures as functions of the B and O chemical potentials, and the corresponding phase diagram. The top panel reports the free-energy hyperplanes corresponding to the most stable phases found for the different stoichiometries we have explored. The lower panel shows the resulting stability phase diagram.\\
The structure characterized by the adsorption sequence Al$_{sub}$- O$_1$ - B$_1$ (c-3 in Fig. \ref{fig:p-all}), is the dominant structure at low B chemical potential (low pressure) and any O pressure. An increase of the B chemical potential determines the transition to stuctures with higher B content, following the sequence AlO$_{\sfrac{1}{2}}$ - B$_2$ (a-1),  AlO$_{\sfrac{3}{4}}$ - B$_2$ (a-2), and AlO$_{1}$ - B$_2$ (a-3)  (see Fig. \ref{fig:p-all}) depending on the oxygen partial pressure. For low O partial pressure and intermediate B chemical potential a stability region for the Al$_{sub}$ - O$_{\sfrac{3}{4}}$ - B$_{\sfrac{5}{4}}$ structure (c-2 in Fig. \ref{fig:p-all}) is present. In the presence of several co-adsorbed species, and with the variety of structures with similar energies that we have identified, these findings should be taken as only qualitative indications of the complexity of the structural landscape accessible in realistic conditions. The actual structures obtained in the lab likely depend on the growth protocol and associated kinetic effects, no less than on thermodynamic stability.

\section{Conclusions \label{sec:conc}}
By combining ab-initio methods and structure prediction algorithms, we have performed an extensive search for the most energetically stable borophene polymorphs on Al(111) substrate to gain insight into the role of electron-deficiency compensation through both self- and charge-doping processes. This was the starting point to investigate geometric- and electronic- structure modifications induced by oxidation as an alternative route to tune the coupling between borophene and the substrate, with respect to self-doping. We conclude that oxygen, due to its high affinity with aluminum, yields passivation of the metal surface and tunes the charge transfer from the metal to borophene. The formation of B-O bonds is hardly observed in the decoupled borophene layers, mostly showing triangular bonding geometries close to the $\alpha$, $\chi_3$, and $\beta_{12}$ structures rather than to the honeycomb layout. Since oxygen hinders the charge transfer from Al to borophene, the resulting electron deficiency is compensated by the formation of other bonding geometries with a higher average filling factor and pronounced buckling. Thus, our results reasonably put in evidence that, in the view of decoupling and stabilizing a honeycosmb borophene phase by means of oxidation, the bonding energy between boron and metal substrate, oxygen and boron, and oxygen and metal substrate should be carefully balanced by a proper choice of the growth substrate. This may be accomplished e.g. by diluting Al and choosing as supporting template an Al alloy with reduced Al-O interaction but, still, with sufficient available charge transfer attitude.
Finally, while our arguments are based on energetic considerations, it should be kept in mind that kinetic effects including surface diffusion, segregation, B gas temperature and pressure affect and actually determine the effective experimental accessibility and observation of each proposed allotropic phase. Thus, our results represent a starting point opening to the first challenging experimental investigations of hB oxidation.\\
\begin{acknowledgments}
MS wishes to thank Yusuf Shaidu and Pietro Delugas for many useful discussions. This work was partially funded by the EU through the \textsc{MaX} Centre of Excellence for supercomputing applications (grant number 824143) and the Italian MUR, through the PRIN project \emph{FERMAT} (grant No. 2017KFY7XF).
\end{acknowledgments}

\appendix

\section{Computational details\label{app:comp}}
All of our calculations have been performed within density-functional theory and the plane-wave pseudopotential methods, as described In Sec. \ref{app:comp-dft} below. Global structural search has been performed using genetic optimization algorithms, as described in Sec. \ref{app:comp-uspex}

\subsection{Plane-wave pseudopotential techniques} \label{app:comp-dft}
We used accurate projector augmented wave method (PAW) pseudo-potentials as recommended in Ref. \onlinecite{sssp} and provided in Ref. \onlinecite{pps}. For the XC functional we have used the generalized gradient approximation exchange-correlation functional parameterized by Perdew-Burke-Ernzerhof (GGA-PBE) \cite{pbe} supplemented by van der waals correction as proposed by Grimme and colleagues \cite{vdw}. The kinetic energy cutoff for charge density is 350 Ry, and energy cutoff for wavefunctions equals 70 Ry, to reach the accurate results . In order to avoid possible inaccuracy due to unwanted dipole field in metal surface, we considered the dipole correction introduced by L. Bengtsson \cite{dip}. On the other hand, the metallic nature of borophene and aluminum requires partial occupation of the energy levels around the Fermi level; we used Marzari-Vanderbilt-DeVita-Payne cold smearing function \cite{marzari} with a line-width $\sigma =$ 0.01 Ry and $(14 \times 14 \times 1)$ Monkhorst-Pack grid for the Brilloun zone integration in the $(2 \times 2)$ surface unit cell. To reach accurate results, optimization of parameters has been considered carefully. The lattice parameter of the substrate is fixed at the bulk value of 4.038 \AA, which is in good agreement with previous works and experimental data \cite{B9n}.

\subsection{Genetic optimization} \label{app:comp-uspex}

The sophisticated genetic methods implemented in USPEX \cite{uspex} help us to investigate overlayer/surface phase space and all the possibilities of their combinations. The algorithm used in USPEX is based on different variation operators, namely, heredity, mutation, and permutation. Briefly, in a first stage, several configurations are generated based on symmetry considerations of crystallography principles. These structures will be considered as the starting point for a fully ab-initio structural optimization, as implemented in \qe\ \cite{QE}, to determine the most energetically favorable configurations accessible in the first generation. These optimized structures will be considered as parents for the next generation of structures, via heredity, mutation, and permutation, followed by ab-initio optimization. This repeated process allows us to find out the most stable and promising structures. In order to access a large variety of the possible configurations, we considered different ratio of boron (oxygen) with respect to aluminum surface.

\bibliography{apsmain}

\begin{thebibliography}{43}
\bibitem [{gro()}]{growth}%
  \BibitemOpen
  \href@noop {} {}\bibinfo {note} {The grwoth temperature is of the order of
  500 K and the M flux is $\approx 0.1$ Ml/min, roughly corresponding to a
  pressure $\lesssim 10^{-8}$ mbar.}\BibitemShut {Stop}%
\end{thebibliography}

\end{document}